\documentstyle{laa}

\begin{document}

\thesaurus{02(12.03.1; 12.12.1)}

\title{Testing the effect of geodesic mixing with COBE data to Reveal
the Curvature of the Universe}

\author{V.G.~Gurzadyan \inst{1} \and S.~Torres \inst{2}}

\offprints{G.~Gurzadyan}

\institute{University of Sussex, Brighton, UK;
Dept. of Theoretical Physics, Yerevan Physics Institute, 375036,
Yerevan, Armenia (permanent address)
\and Observatorio Astron\'omico Nacional, Universidad Nacional de Colombia,
and Centro Internacional de F\'{\i}sica,
Bogot\'a, Colombia}

\date{Received ; accepted }

\maketitle

\begin{abstract}
When considering the statistical properties
of a bundle of cosmic microwave background (CMB)
photons propagating through space,
the effect of `mixing geodesics' appears with a distinct signature that
depends on the geometry of space.
In a Universe with negative curvature this effect
is expected to produce elongated anisotropy spots on CMB maps.
We used COBE-DMR data to look for such effect. Based on
the analysis of a measure of eccentricity of hot spots
it appears that there is a clear indication of an excess
eccentricity of hot spots with respect to that expected from
noise alone. This result must be interpreted with 
caution as this effect can be due in part to 
galactic emission. 
If the detected eccentricity of anisotropy spots
can be attributed to the effect of mixing it implies
the negative curvature of the Universe and a value of $\Omega < 1$.
\end{abstract}

\section{Introduction}

The detection (Smoot et al. 1992; Bennett et al. 1994)
and further confirmation (Bennett et al. 1996;
Hancock et al. 1994; Ganga et al. 1993) of the presence of
anisotropies in the cosmic microwave background (CMB),
has endowed cosmologists with a unique tool for
a realistic test of cosmological models.
The observed large angle $\Delta T/T$
can be produced by a number
of sources that go from local, to astrophysical, to cosmological
effects, among which the most relevant are: emission from the
Galaxy;
gravitational
fluctuations produced by voids and great attractors;
the Doppler
effect due to the observer's motion relative to the CMB;
mixing of geodesic flow;
gravitational waves;
cosmic strings; and
gravitational potential fluctuations on the
last scattering surface --
the Sachs-Wolfe effect (for a review see
White, Scott \& Silk 1994). Clearly the
complexity of determining the cause of CMB
anisotropies requires several independent analysis techniques.
Aside from the standard statistical tests based
on the angular auto-correlation function and power
spectrum,
a number of statistics have been
proposed (Sazhin 1985;
Vittorio \& Juszkiewicz 1987;
Bond \& Efstathiou 1987;
Coles 1988;
Gott et al. 1990;
Mart\'\i nez-Gonz\'alez \& Cay\'on 1992;
Mart\'\i nez-Gonz\'alez \& Sanz 1989)
as topological descriptors to characterize
the observed anisotropies:
hot spot number density, length and
total curvature or {\it genus} of iso-temperature contours,
number of upcrossings,
area and eccentricity
of hot spots, and Euler-Poincar\'{e} characteristic.
The study of these topological descriptors
is based on the well established theory of the  geometric
properties of the excursion set
of random fields (Adler 1981).

The great potential of topological analysis of
CMB data can be attested by the large number of
results obtained thus far:
the {\it genus} descriptor has been successfully used
to place important restrictions
on the shape of the spectrum of primordial
perturbations $P(k)$ (Smoot et al. 1994; Torres 1994; 
Torres 1995a; Torres et al. 1995);
further restrictions on $P(k)$ have been obtained from
the peak distribution analysis (Fabbri \& Torres 1995, 1996);
a possible fractal dimension in the universe has
been detected with the analysis of the contour
characteristics of anisotropy spots (De Gouveia 1995);
studies of hot spot number density (Torres 1995b; Cay\'on \&
Smoot 1995)
support the results given by the {\it genus} analysis;
tests of Gaussianity of the CMB field have been
performed using
the genus and the correlation function of local
maxima (Kogut et al. 1996);
finally,
percolation and cluster analysis (Naselsky \& Novikov 1995) of CMB
maps has been proposed as a powerful tool to test
the Gaussian character of CMB anisotropies. The information
carried by the CMB properties can even have a direct 
impact on accelerator
physics (Gurzadyan \& Margarian 1996).

We propose to use a topological analysis, based on the
eccentricity or elongation measure
of hot spots, to look for the
signature of the mixing of geodesics effect 
(Gurzadyan \& Kocharyan 1991, 1992).
Because of the appealing possibility of extracting
information about the curvature of the universe from the latter
effect, we will deal with it in some detail.
A consequence of mixing  phenomena in geodesics propagation
is the appearance of highly distorted anisotropy spots in
cosmic background maps (Gurzadyan \& Kocharyan 1993a, 1993b).
In the case of an open geometry,
anisotropy spots
associated with a bundle of photons propagating in free
space would appear as highly elongated hot spots,
while in a flat space hot spots would show a
symmetric circular shape.
This visible effect is a direct consequence
of the geometry of space and as such it is a unique
tool to discriminate between open and closed cosmological
models.
However, confusion due
to galactic contamination at high latitudes and similar elongated
patterns produced by instrumental noise and
cosmological structure characterized by large coherence angles
makes this effect
very difficult to observe.

Fortunately, at least 3 different phenomena produced by
the effect of mixing geodesics and associated with the observable
characteristics of CMB, have been predicted:

(a) {\it Isotropization.} The decrease of the amplitude 
of CMB anisotropy
after the decoupling epoch: the degree of isotropization, i.e. 
the fluctuation
damping factor can yield up to $10^3$ for $\Omega$ = 0.2-0.4
(Gurzadyan \& Kocharyan 1991,1992)\footnote{In Eq(12) of the interesting
paper by Ellis and Tavakol (1994) an  additional numerical factor 2 is
necessary, as readily follows from Pesin's theorem; in the rest those
authors indeed confirm the above mentioned results on the decrease of
CMB anisotropy via geodesic mixing.};

(b) {\it Angular dependence.} The behaviour of anisotropy as a function of
the sky and smoothing (beam)
angles, namely, the independence of the autocorrelation function on the sky
angle, and its dependence on the smoothing angle 
(Gurzadyan \& Kocharyan 1996);

(c) {\it Maps.} The complex topological structure of the CMB sky maps,
showing intrinsic {\it threshold independent} elongation
of the shapes of both hot and cold spots 
(Gurzadyan \& Kocharyan 1993a, 1993b).

In this paper we will deal only with the third phenomenon.
Since, in principle, effects of apparent elongation
might occur also owing to other physical mechanisms (the same is not excluded
also for phenomena (a) and (b)), the unambiguous confirmation
of the observational discovery of the effect of mixing geodesics could
be reached by means of a comprehensive study of the
complete set of predictions. As it is
briefly discussed at the end of this paper, at least the available evidences
do not contradict to all these predictions.
A confirmation of the effect of mixing geodesics would
constitute powerful probe of the curvature of the Universe.

\section{Effect of mixing of geodesics flow: Maps}

The idea of the effect of instability of trajectories of freely
moving particles can be
most clearly demonstrated via the Jacobi equation  written for spaces
with constant curvature $k$:
\begin{equation}
d^2 {\bf n}/d^2 \lambda + k{\bf n} = 0,
\end{equation}
describing the behavior of the vector of deviation, {\bf n},
of close geodesics.
Solutions of this equation are determined by the sign of
the curvature: when $k<0$ one has exponentially deviating geodesics.

A more rigorous treatment
(Gurzadyan \& Kocharyan 1991, 1992)
includes the study of the projection of geodesics from
(3+1)-dimensional Lorentzian space to a 3D Riemannian one, and
the behavior of time correlation functions for
geodesic flows on homogeneous isotropic  spaces with negative curvature.
Geodesic flows, being Anosov systems
(locally if the space is not compact),
are exponentially unstable systems
possessing: the strongest statistical properties (mixing),
non-zero Lyapunov
characteristic exponents, and
positive Kolmogorov-Sinai (KS) entropy~$h$ (Arnold 1989).

For Anosov systems two geodesics in 3-space deviate exponentially
according to the law
\begin{equation}
{\bf n}(\lambda) = {\bf n}(0)\exp (\chi \lambda),
\end{equation}
where $\chi$ is the Lyapunov exponent.

For a  homogeneous isotropic Friedmannian Universe with $k=-1$ the
Lyapunov exponent
is determined by the only parameter $a$, the diameter of the Universe:
\begin{equation}
\chi=1/a,
\end{equation}
while Lyapunov exponents vanish when $k=0,+1$.

Time correlation functions,  describing the decrease of perturbations
also decay exponentially as  determined by KS-entropy:
$h=2\chi$.

For the Universe expanding as
\begin{equation}
a(t)=a(t_0)(t/t_0)^{\alpha},
\end{equation}
the relation between the quantitative measurement of
the distortion  of patterns,
$\epsilon$, and $\Omega$ is given by
(Gurzadyan \& Kocharyan 1991, 1992):

\begin{equation}
\frac{\ln {(1/\epsilon)}}{(\frac{1-\Omega_0}{1+z\Omega_0})^{\frac{1}{2}}} =
\left\{ \begin{array}{ll}
 \alpha/(1-\alpha) [1-(1+z_1)^{1-1/\alpha}] & \alpha<1\\
                                  \ln(1+z_1) &  \alpha=1
 \end{array}
 \right.
\end{equation}
where $\Omega_{0}$ is its present value ($z=0$),
$z_1$ corresponds to the time when matter becomes non relativistic and
$z (\approx 1000)$ to the decoupling time.
We deliberately used the power law representation for the expansion law,
(instead of trivial use of the corresponding exact Einstein
equation) to demonstrate the role of the expansion rate
in the described effect. It is not excluded that in the future
the geodesic mixing effect can be useful to gain information also on the
expansion rate of the Universe after the decoupling epoch.

The parameter of elongation $\epsilon$  defined via the divergence
of geodesics in (3+1)-space:
\begin{equation}
\epsilon=\frac{l(t)}{l(0)}, \
l(t)=l(t_0)\frac{a(t)}{a(t_0)} \exp (\chi \lambda(t)),
\end{equation}
approaches 1 when $\Omega$ tends to 1 as shown in Fig.~1.

\begin{figure}
\vspace{4.5cm}
\caption[]{Elongation parameter $\epsilon$ as a function of $\Omega$
for two exponents for the law of the expansion rate of the Universe,
$\alpha = 1$ and $\alpha = 2/3$.}
\end{figure}

The typical pattern of
a hot spot as seen today in a $k<1$ universe would exhibit a very
complex shape.
The elongation, $\epsilon$,
measures the
smallest-to-largest ratio of diameters of one-connected regions and
is related to the
``degree of complexity'' of anisotropies.

\section{Analysis of DMR Maps}

COBE's Differential Microwave Radiometers (DMR) have mapped the
microwave sky
at three frequencies: 31, 53, and 90 GHz (Smoot et al. 1990).
The analysis presented here is based on DMR's 53 GHz
maps (which have the best sensitivity) of the
four year data set (Bennett et al. 1996), and is a
continuation of the study based
on the one year data-set (Gurzadyan \& Torres 1993).
Signal and noise maps were prepared by
adding and subtracting the two independent DMR channels
(ie. $0.5A + 0.5B$ and $0.5A - 0.5B$)
and Gaussian smoothing ($\sigma = 2.9^{\circ}$) the resulting
maps.
A Galaxy cut $|b| < 20^{\circ}$ has been applied to the maps.
The geometric characteristics of hot spots are quite sensitive
to galactic cuts below  $15^{\circ}$ to $20^{\circ}$ but
beyond $20^{\circ}$ our results are stable.

DMR maps are digitized on a grid of 6144
approximately equal area pixels of size
$\sim 2.9^{\circ}$ (Torres et al.  1989).
A hot spot is defined as a continuous
region of the map formed by pixels whose temperature is
higher than a preset threshold $T_{\nu} = \nu \sigma$,
where $\sigma$ is the standard deviation of the sky temperature
(monopole and dipole subtracted).
The algorithm to identify hot spots (Torres 1994, Torres 1995a)
relies on the construction of binary tree structures out of
the set of connected `hot' pixels (for a given temperature
threshold). The number of pixels in a tree gives the area
of the spot.
The  eccentricity parameter
$\epsilon^i_{\nu} \equiv r^i_{min}/r^i_{max}$ of the
$i$-th hot spot at threshold $\nu$ is found as follows:
first, the center of the spot is estimated as the
point with $x_c,y_c$ coordinates equal to the
mean of all the $x$ and $y$ coordinates
of the pixels that form the spot respectively;
second, the distance from this center point to all the pixels
that lay on the boundary of the spot are calculated;
$r^i_{max}$ and $r^i_{min}$ are the largest
and shortest of these distances.
The above mentioned procedure is repeated
for several threshold levels ($\nu: 1.0 - 3.0$ in
steps of $\Delta \nu = 0.25$).
The geometric descriptor $\epsilon_{\nu}$ used to measure the
elongation of anisotropy spots at threshold $\nu$
is the average of all the $\epsilon^i_{\nu}$ at a fixed
$\nu$.

At this point one can question whether the defined
$\epsilon_{\nu}$ really
corresponds to the theoretical parameter of elongation in Fig.~1.
The correspondence of the Lyapunov numbers or KS-entropy with
parameters observed either with experiments or computer simulations
is an essential problem. However, it has been empirically
established that
a correspondence
with macroscopic parameters does exist (at least qualitative)
and just this fact determines
the role of those parameters
as important tools for the study of nonlinear phenomena
(Hilborn 1994 and references therein).
Another question is whether the specific procedure
of estimation of the mean elongation of spots can be crucial or not.
Obviously other algorithms in principle can lead to numerically
not absolutely equivalent results, but a drastic
change of the situation seems
unlikely, especially at high thresholds (Sommerer \& Ott 1993 and references
therein).

\begin{figure}
\vspace{4.5cm}
\caption[]{Measured eccentricity parameter $\epsilon_{\nu}$
for the COBE sum maps ({\it solid}) and difference maps
({\it dash}) compared with the expected eccentricity for
noise Monte Carlo realizations ({\it points with error bars}).
}
\end{figure}

Figure 2 shows
$\epsilon_{\nu}$ for the sum and difference DMR maps
for threshold levels in the range $\nu: 1.0 - 3.0$.
Data for $\nu < 1.0$ is dominated by noise and
for $\nu > 3.0$ is limited by the large statistical
error due to the small number of spots at high thresholds.

\section{Results}

The signature of the mixing of geodesics effect
is a clear one: hot spots should have a fixed
eccentricity independent of threshold level, and
if these hot spots are elongated it is an indication
of mixing in open spaces. In principle it is possible
to detect eccentricities smaller than 1, the problem
however is to disentangle the effect due to mixing
from instrumental noise which by its stochastic
nature is expected to produce anisotropy spots
with certain degree of elongation.

In order to evaluate the statistical significance
of a possible detection of the mixing of geodesics effect
we have performed Monte Carlo studies of noise maps
that take into account instrumental noise,  COBE's
sky coverage and pixelization.
Noise maps were generated by assigning to each pixel
on the map a temperature equal to a random number extracted
from a Gaussian distribution with dispersion
$\sigma = \sigma_{1}/\sqrt N$, where
$\sigma_{1}$ is the corresponding sensitivity for
one observation and $N$ the number of observations.

Maps for both $A$ and $B$ channels were generated
independently. A difference map was formed and
Gaussian smoothed just as it is done with the real data.
The same algorithm used to obtain the eccentricity
parameter of the COBE maps was used for each one of
the Monte Carlo noise realizations. Figure 2
shows the Monte Carlo mean eccentricity and the 1-$\sigma$ error
bars expected from noise maps.
Knowing the eccentricity expected
from noise, one can estimate the probability
that the observed eccentricity parameter can be
produced by noise alone. Table 1 gives $\epsilon_{\nu}$
for the COBE sum and difference maps, and the
deviations from the mean eccentricity of
Monte Carlo noise simulations.

%\begin {center}
\begin{table}
\caption{Eccentricity parameter of hot spots on COBE maps
($\epsilon_{\nu}^{A+B}$, $\epsilon_{\nu}^{A-B}$)
and comparison with Monte Carlo noise maps
($\epsilon_{\nu}^{MC}$).
$\Delta^{A+B}$ and $\Delta^{A-B}$ denote
the difference (in standard deviations)
between the measured eccentricities
and the mean eccentricity of
noise Monte Carlo maps.}
\vspace{4mm}
\begin{center}
%\begin{tabular}{l@{\quad}l@{\quad}l@{\quad}l@{\quad}l@{\quad}l@{\quad}l}
\begin{tabular}{rrrrrrr}
\hline\hline \\
$\nu$ &
$\epsilon_{\nu}^{A+B}$ &
$\epsilon_{\nu}^{A-B}$ &
$\epsilon_{\nu}^{MC}$ &
$\Delta^{A+B}$ &
$\Delta^{A-B}$ \\[2pt]
\hline\rule{0pt}{12pt} \\
1.00 &    0.480 &    0.570 &    0.582  &   3.551 &    0.447 \\
1.25 &    0.547 &    0.590 &    0.617  &   2.271 &    0.886 \\
1.50 &    0.497 &    0.616 &    0.646  &   4.288 &    0.881 \\
1.75 &    0.576 &    0.691 &    0.674  &   2.432 &   -0.429 \\
2.00 &    0.600 &    0.686 &    0.699  &   2.049 &    0.267 \\
2.25 &    0.595 &    0.819 &    0.719  &   2.060 &   -1.675 \\
2.50 &    0.427 &    0.685 &    0.741  &   4.017 &    0.719 \\
2.75 &    0.574 &    0.732 &    0.761  &   1.752 &    0.272 \\
3.00 &    0.507 &    0.705 &    0.776  &   1.815 &    0.480 \\ [2pt]
\hline
\end{tabular}
\end{center}
\end{table}
%\end{center}

The $\chi^2$ statistic computed with the 9 data points
in the range $\nu = 1.0 - 3.0$ and the corresponding
noise Monte Carlo points is 5.6 and 73.0 for the $(A-B)$ and
$(A+B)$ maps respectively. The low $\chi^2$
associated with the difference maps was expected
and is an indication of the accuracy of
the Monte Carlo simulations.
On the other hand, the high $\chi^2$ obtained
when the data from the signal maps is compared with
noise data is a clear indication of an actual detection
of elongated anisotropy spots. The measured eccentricity
parameter at threshold levels 1.5 and 2.5 in particular
exhibit a $>4$-$\sigma$ deviation with respect to noise.
The average deviation in terms of standard deviations
of $\epsilon_{\nu}^{A+B}$ from
the corresponding Monte Carlo result for the
9 bins considered here is $3\sigma$.
From the mixing of geodesics effect one
would expect a constant eccentricity for all threshold
levels. Due to the large dispersion of the measured
$\epsilon_{\nu}$ values it is not possible to
make a strong statement in favor of the hypothesis
of a constant $\epsilon_{\nu}$ independent of $\nu$. However,
it is seen that the contribution to the $\chi^2$ is
roughly the same independent of $\nu$.  Under the
hypothesis of a positive detection of elongated
anisotropy spots and using the 9 points in Table 1,
the measured eccentricity is $\epsilon = 0.53 \pm 0.06$.

\section{Discussion}

An excess elongation as possible genuine feature of the
hot spots on COBE maps is well established at least
at threshold levels $\nu$ between 1.0 and 3.0. The
detected excess elongation shows a statistically strong
independence on the threshold, contrary to the case for the
difference $(A-B)$ maps where $\epsilon_{\nu}$ shows a 
clear correlation
with threshold. This characteristic dependence of the eccentricity
parameter with threshold for noise maps is
clearly manifest in Monte Carlo noise realizations.
If the detection of high eccentricity of hot spots
is attributed to the effect of mixing of geodesics flow,
our analysis implies  $\Omega<1$. However one must also consider 
the fact that the detected effect can be produced in part by 
galactic contamination 
of the maps. 

The
numerical value of the obtained elongation parameter might have been
obviously affected by a number of effects of cosmological and
non-cosmological nature.
Primordial fluctuations on the surface of last
scattering, for example, would also produce
non-zero ellipticity even in flat universe
(Bond \& Efstathiou 1987).
However, due to the stochastic nature of this effect,
the dependence of the ellipticity parameter with threshold
would be just like that of noise, contrary to what
we have observed (i.e. a threshold independent
eccentricity).
More statistics and higher accuracy is
necessary for a reliable numerical evaluation of $\Omega$.
A question thus arises as to how one can distinguish whether the
observed elongation
is really due to the effect of geodesics mixing?
Indeed, although the ``geodesics mixing'' has put
forward the idea of looking
for the {\it threshold independent} elongation of hot spots on CMB sky maps
as a signature for $\Omega <1$,
there are other mentioned effects that can be considered
as well. Thus,
to answer the question posed above it is important to
have a thorough
consideration of other observable consequences of
the effect of mixing geodesics.
As it is mentioned above, among those consequences is the decrease of the
amplitude of initial fluctuations during the
expansion of the Universe; the expected decrease depends on
the curvature, the value of $\Omega$ and the time of last scattering
(Gurzadyan \& Kocharyan 1992).

More interesting, however, can be the property concerning  the
dependence of the temperature autocorrelation function $C(\theta,\beta)$
on the sky angle $\theta$ with a beam size $\beta$.
It was shown (Gurzadyan \& Kocharyan 1996), 
that in negatively curved spaces  $C(\theta,\beta)$
tends to become constant with respect to $\theta$ (i.e. independent on the
sky angle) independent of the initial perturbation spectrum,
but should depend on the beam size $\beta$ by the asymptotic law
$$
\frac{C(\theta,\beta)}{\bar{T^2}} -1 \sim \frac{\rm const}{\beta}.
$$
In other words, the observation with different beam sizes should lead to
different values for the amplitude of the CMB anisotropy, namely the
measured anisotropy decreases with increasing beam size.

Regarding the latter effect, note that some
measurements by smaller
beam sizes (Netterfield et al 1995; Ruhl et al 1995) seem to indicate higher
values for the anisotropy amplitude as compared with the COBE data,
even though
one has to distinguish the contributions of other effects
such as the Doppler peak
for example.

The peculiar feature of the effect of geodesic is its crucial dependence
on the curvature of the Universe, i.e. its
disappearance in flat and positively
curved spaces, and independence on various models of dark matter or
the spectrum of initial perturbations, etc.

Obviously nothing is happening to an individual photon
during the free propagation
after the last scattering epoch,
and these effects are entirely statistical and
determined by the principal limitations
of obtaining information during
the measurements, i.e. by the impossibility of the reconstruction of the
trajectory of the photon while observing within finite smoothing angle
and time period, due to the overlapping of
exponentially deviating geodesics in any given
cut of phase space (for detailed discussion
of this and similar problems see Gurzadyan \& Kocharyan 1994). Note that
these effects differ from those of classical chaotic Mixmaster models,
and are alternatives to inflationary scenarios
(Gurzadyan \& Kocharyan 1994; Cornish et al 1996).
The measured effect of elongation of anisotropy spots
in CMB sky maps can be the
direct footprint of the negative curvature of the Universe and
therefore has
a considerable impact on cosmology.

\begin{acknowledgements}
The authors
would like to thank R.Juszkiewicz, A.Kocharyan and R.Ruffini for
valuable discussions of the results.
One of us (V.G.) is thankful to R.Penrose
for the discussion of the geometrical aspect of the effect of
geodesics mixing. This research is in
part funded by COLCIENCIAS of Colombia (2228-05-091-95) and The Royal Society.
The COBE datasets were developed by the NASA Goddard
Space flight Center under the guidance of the COBE Science
Working Group and were provided by the NSSDC.
\end{acknowledgements}


\begin{thebibliography}{99}

\bibitem{Adler} Adler, R. J. 1981, {\it The Geometry of Random Fields},
(Chichester: Wiley)

\bibitem{arnold} Arnold, V.I. 1989, {\it Mathematical Methods of Classical
Mechanics}, Nauka, Moscow

\bibitem{bennett94} Bennett, C. L., et al.1994, ApJ, 436, 423

\bibitem{bennett96} Bennett, C. L., et al. 1996, ApJ, 464, L1

\bibitem{bond} Bond, J.R. \& Efstathiou, G. 1987, MNRAS,
226, 655

\bibitem{cayon95} Cay\'{o}n, L. \& Smoot, G. F. 1995, ApJ, 452, 487


\bibitem{Coles} Coles, P. 1988, MNRAS, 234, 509

\bibitem{Corn} Cornish N.J., Spergel D.N. \& Starkman G.D. 1996
astro-ph/9601034

\bibitem{thyrso} De Gouveia, E. M., et al. 1995, ApJ, 442, L45

\bibitem{ellis} Ellis G., Tavakol R. 1994, Class.Quant.Grav., 11, 675 

\bibitem{fabbri95} Fabbri, R. \& Torres, S. 1995, Nuovo Cimento, 110B, 865

\bibitem{fabbri96} Fabbri, R. \& Torres, S. 1996, A\&A, 307, 703

\bibitem{cheng} Ganga, K., Cheng, E., Meyer, S. \& Page, L. 1993, ApJ, 410,
L57

\bibitem{Gott} Gott, J. R.  et al. 1990, ApJ, 352, 1

\bibitem{GK1} Gurzadyan V.G. \& Kocharyan, A.A. 1991, in: {\it Quantum Gravity, V},
(eds Berezin V.A., Markov M.A. \& Frolov V.P.) p.689 (World Sci. Singapore).

\bibitem{GK2} Gurzadyan, V.G. \& Kocharyan, A.A. 1992,  A\&A, 260, 14

\bibitem{mixing} Gurzadyan, V.G. \& Kocharyan, A.A. 1993a, Int. Journ. 
Mod. Phys. D., 2, 97

\bibitem{mix2} Gurzadyan, V.G. \& Kocharyan, A.A. 1993b, Europhys. Lett. 22, 231

\bibitem{GuKo} Gurzadyan V.G. \& Kocharyan, A.A. 1996,
 in: {\it Quantum Gravity, VI}, (eds Berezin V.A., \& Rubakov V.A.)
(World Sci. Singapore, in press); ICTP Preprint IC/93/419, Trieste, 1993

\bibitem{gkbook} Gurzadyan, V.G. \& Kocharyan, A.A. 1994, {\it Paradigms of the
Large-Scale Universe}, (Gordon and Breach, New York).


\bibitem{GuM} Gurzadyan, V.G., \& Margarian, A. 1996, Physica Scripta, 53, 513

\bibitem{GuTorres} Gurzadyan, V.G. \& Torres, S. 1993,
in: {\it Present and Future of the Cosmic Microwave
Background}, (Eds. J.L. Sanz, et al.) 429, p.139 (Heidelberg: Springer-Verlag)

\bibitem{hancock} Hancock, S., et al. 1994, Nature, 367, 333

\bibitem{Hi} Hilborn, R.C. 1994, {\it Chaos and Nonlinear Dynamics}, (Oxford
University Press, Oxford).

\bibitem{kogut96} Kogut, A., et al. 1996, ApJ, 464 L5

\bibitem{Martinez} Mart\'{\i}nez-Gonz\'{a}lez, E., Cay\'on, L. 1992, in
{\it The Infrared and Submillimetre Sky after COBE}
(eds M. Signore and C. Dupraz) 303 (Netherlands: Kluwer Academic Press)

\bibitem{MarSanz} Mart\'{\i}nez-Gonz\'{a}lez, E. \& Sanz, J. L. 1989, MNRAS,
237, 939

\bibitem{novikov} Naselsky, P. D., \& Novikov, D. I. 1995, ApJ, 444, L1

\bibitem{Netter} Netterfield, C.B., et al. 1995, ApJ, 445, L69

\bibitem{Ruhl} Ruhl J.E., et al. 1995, ApJ, 453, L1

\bibitem{Sazhin}  Sazhin, M. V. 1985, MNRAS, 216, 25p

\bibitem{Smoot90} Smoot, G.F., et al. 1990, ApJ, 360, 685

\bibitem{Smoot92} Smoot, G. F., et al.1992,  ApJ, 396, L1

\bibitem{smoot95} Smoot, G. F., et al. 1994, ApJ, 437, 1

\bibitem{So} Sommerer, J.C., \& Ott, E. 1993, Science, 259, 335

\bibitem{torres89} Torres, S., et al. 1989, in {\it Data Analysis in Astronomy III}
(eds V. di Gesu, L. Scarsi, and  M.C. Maccarone) 319  (New York: Plenum)

\bibitem{torres94a} Torres, S. 1994, ApJ, 423, L9

\bibitem{torres95a} Torres, S., et al. 1995, MNRAS, 274, 853

\bibitem{torres95a} Torres, S. 1995a, Astro. Lett. and Communications,
32, 95

\bibitem{torres94c} Torres, S. 1995b, Astrophys. and Space Sci. 228, 313


\bibitem{Vittorio} Vittorio, N. \& Juszkiewicz, R. 1987,  ApJ, 314, L29

\bibitem{WSS} White, M., Scott, D. \& Silk, J. 1994, Ann.Rev.A \& A, 32, 319

\end{thebibliography}
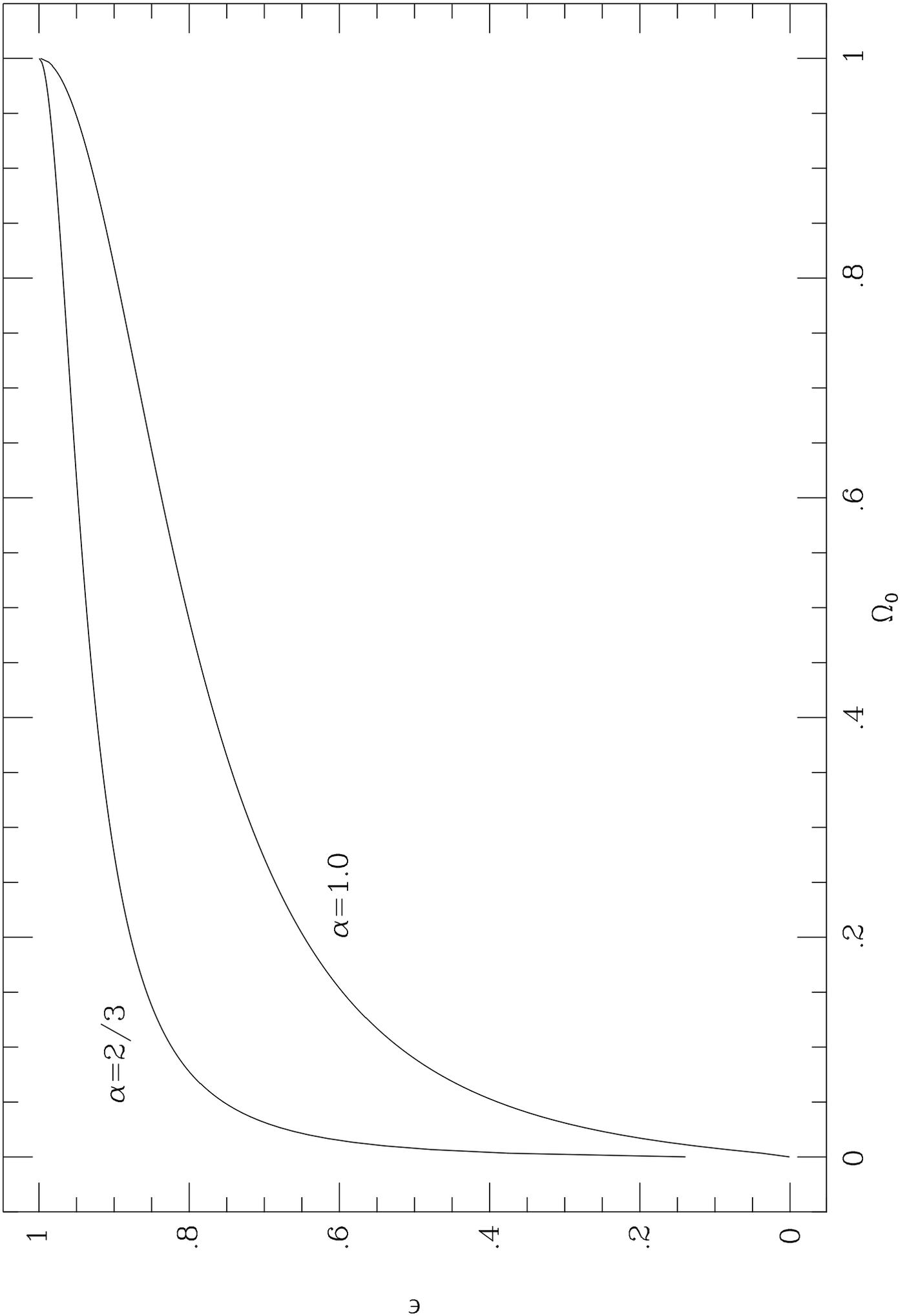
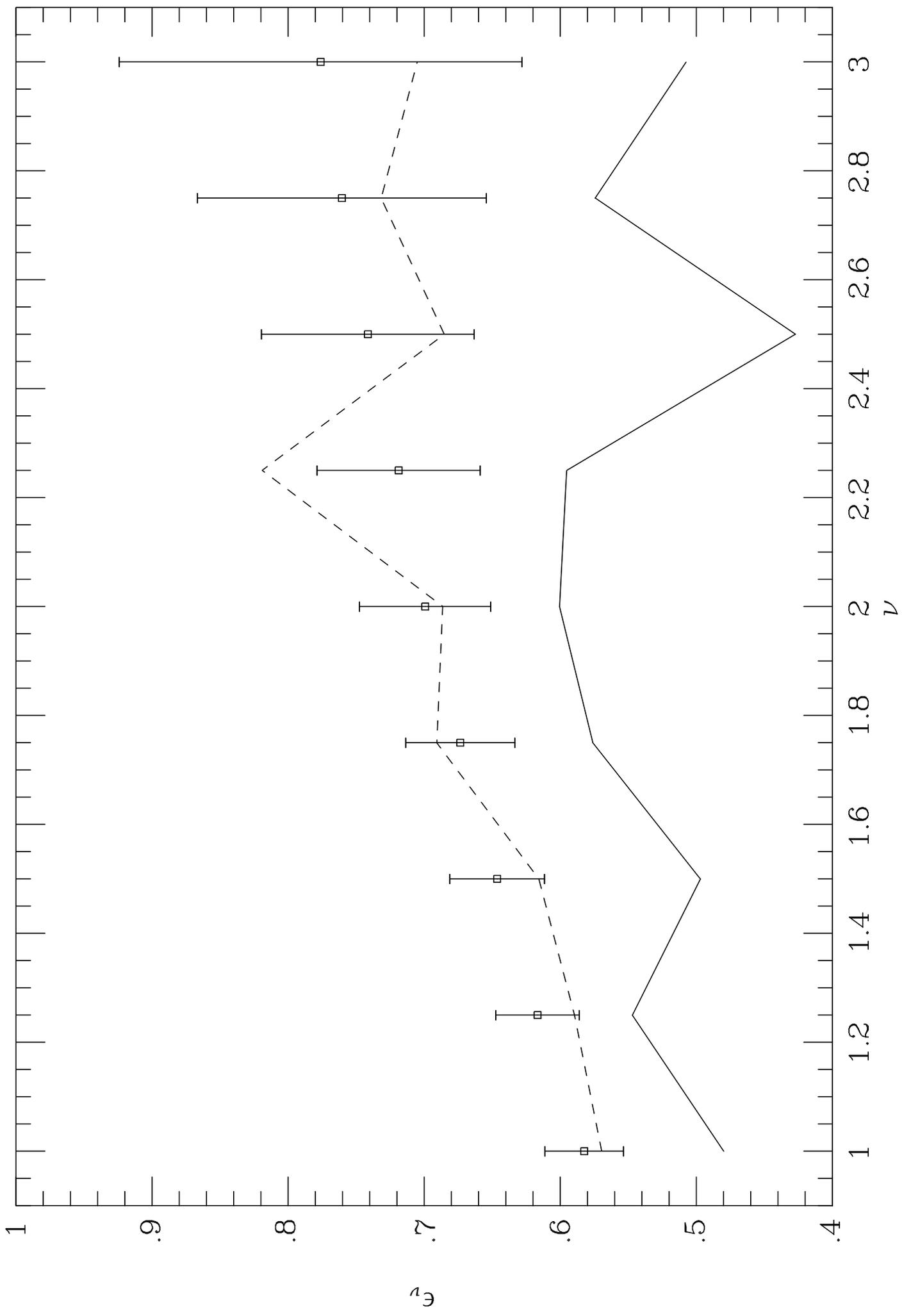
\end{document}